\pdfoutput=1

\documentclass[10pt,a4paper,twocolumn]{article}

\usepackage{ifthen} 
\newboolean{pdflatex}
\setboolean{pdflatex}{true} 

\newboolean{articletitles}
\setboolean{articletitles}{true} 

\newboolean{uprightparticles}
\setboolean{uprightparticles}{false} 

\newboolean{inbibliography}
\setboolean{inbibliography}{false} 

\usepackage{microtype}
\usepackage{lineno}  
\usepackage{xspace} 
\usepackage{caption} 

\usepackage{graphicx}  
\usepackage{color}
\usepackage{colortbl}
\graphicspath{{./}} 

\usepackage{amsmath} 
\usepackage{amssymb}
\usepackage{amsfonts}
\usepackage{upgreek} 

\newcommand*\patchAmsMathEnvironmentForLineno[1]{%
\expandafter\let\csname old#1\expandafter\endcsname\csname #1\endcsname
\expandafter\let\csname oldend#1\expandafter\endcsname\csname
end#1\endcsname
 \renewenvironment{#1}%
   {\linenomath\csname old#1\endcsname}%
   {\csname oldend#1\endcsname\endlinenomath}%
}
\newcommand*\patchBothAmsMathEnvironmentsForLineno[1]{%
  \patchAmsMathEnvironmentForLineno{#1}%
  \patchAmsMathEnvironmentForLineno{#1*}%
}
\AtBeginDocument{%
\patchBothAmsMathEnvironmentsForLineno{equation}%
\patchBothAmsMathEnvironmentsForLineno{align}%
\patchBothAmsMathEnvironmentsForLineno{flalign}%
\patchBothAmsMathEnvironmentsForLineno{alignat}%
\patchBothAmsMathEnvironmentsForLineno{gather}%
\patchBothAmsMathEnvironmentsForLineno{multline}%
\patchBothAmsMathEnvironmentsForLineno{eqnarray}%
}

\usepackage{hyperref}    
\usepackage[all]{hypcap} 

\def\lhcb {\mbox{LHCb}\xspace}

\def\babar  {\mbox{BaBar}\xspace}

\def\MagUp {\mbox{\em Mag\kern -0.05em Up}\xspace}

\ifthenelse{\boolean{uprightparticles}}%
{

 \def\Ppi         {\ensuremath{\uppi}\xspace}                 
                  
 \def\Prho        {\ensuremath{\uprho}\xspace}

 \def\PDelta      {\ensuremath{\Delta}\xspace}                 
 \def\PXi      {\ensuremath{\Xi}\xspace}                 
 \def\PLambda      {\ensuremath{\Lambda}\xspace}                 
 \def\PSigma      {\ensuremath{\Sigma}\xspace}                 
 \def\POmega      {\ensuremath{\Omega}\xspace}                 
 \def\PUpsilon      {\ensuremath{\Upsilon}\xspace}                 
 
 \def\PB      {\ensuremath{\mathrm{B}}\xspace}                 
                  
 \def\PD      {\ensuremath{\mathrm{D}}\xspace}

 \def\PK      {\ensuremath{\mathrm{K}}\xspace}

 \def\Pb      {\ensuremath{\mathrm{b}}\xspace}                 
 \def\Pc      {\ensuremath{\mathrm{c}}\xspace}                 
 \def\Pd      {\ensuremath{\mathrm{d}}\xspace}

 \def\Pi      {\ensuremath{\mathrm{i}}\xspace}

 \def\Pu      {\ensuremath{\mathrm{u}}\xspace}

}
{

 \def\Ppi         {\ensuremath{\pi}\xspace}                 
                  
 \def\Prho        {\ensuremath{\rho}\xspace}

 \mathchardef\PDelta="7101
 \mathchardef\PXi="7104
 \mathchardef\PLambda="7103
 \mathchardef\PSigma="7106
 \mathchardef\POmega="710A
 \mathchardef\PUpsilon="7107
                  
 \def\PB      {\ensuremath{B}\xspace}                 
                  
 \def\PD      {\ensuremath{D}\xspace}

 \def\PK      {\ensuremath{K}\xspace}

 \def\Pb      {\ensuremath{b}\xspace}                 
 \def\Pc      {\ensuremath{c}\xspace}                 
 \def\Pd      {\ensuremath{d}\xspace}

 \def\Pi      {\ensuremath{i}\xspace}

 \def\Pu      {\ensuremath{u}\xspace}

}

\makeatletter
\ifcase \@ptsize \relax
  \newcommand{\miniscule}{\@setfontsize\miniscule{4}{5}}
\or
  \newcommand{\miniscule}{\@setfontsize\miniscule{5}{6}}
\or
  \newcommand{\miniscule}{\@setfontsize\miniscule{5}{6}}
\fi
\makeatother

\DeclareRobustCommand{\optbar}[1]{\shortstack{{\miniscule (\rule[.5ex]{1.25em}{.18mm})}
  \\ [-.7ex] $#1$}}

\def\uquark    {{\ensuremath{\Pu}}\xspace}

\def\dquark    {{\ensuremath{\Pd}}\xspace}
\def\dquarkbar {{\ensuremath{\overline \dquark}}\xspace}

\def\cquark    {{\ensuremath{\Pc}}\xspace}

\def\bquark    {{\ensuremath{\Pb}}\xspace}

\def\pion   {{\ensuremath{\Ppi}}\xspace}
\def\piz    {{\ensuremath{\pion^0}}\xspace}

\def\pip    {{\ensuremath{\pion^+}}\xspace}
\def\pim    {{\ensuremath{\pion^-}}\xspace}
\def\pipm   {{\ensuremath{\pion^\pm}}\xspace}

\def\rhomeson {{\ensuremath{\Prho}}\xspace}
\def\rhoz     {{\ensuremath{\rhomeson^0}}\xspace}
\def\rhop     {{\ensuremath{\rhomeson^+}}\xspace}
\def\rhom     {{\ensuremath{\rhomeson^-}}\xspace}

\def\kaon    {{\ensuremath{\PK}}\xspace}
  \def\Kbar    {{\kern 0.2em\overline{\kern -0.2em \PK}{}}\xspace}

\def\KorKbar    {\kern 0.18em\optbar{\kern -0.18em K}{}\xspace}

\def\Kp      {{\ensuremath{\kaon^+}}\xspace}
\def\Km      {{\ensuremath{\kaon^-}}\xspace}

  \def\Dbar    {{\kern 0.2em\overline{\kern -0.2em \PD}{}}\xspace}
\def\D       {{\ensuremath{\PD}}\xspace}

\def\DorDbar    {\kern 0.18em\optbar{\kern -0.18em D}{}\xspace}
\def\Dz      {{\ensuremath{\D^0}}\xspace}
\def\Dzb     {{\ensuremath{\Dbar{}^0}}\xspace}

\def\Dstarp  {{\ensuremath{\D^{*+}}}\xspace}

\def\Bbar    {{\ensuremath{\kern 0.18em\overline{\kern -0.18em \PB}{}}}\xspace}

\def\BorBbar    {\kern 0.18em\optbar{\kern -0.18em B}{}\xspace}

  \def\Y#1S{\ensuremath{\PUpsilon{(#1S)}}\xspace}

\def\Lbar        {{\ensuremath{\kern 0.1em\overline{\kern -0.1em\PLambda}}}\xspace}
\def\LorLbar    {\kern 0.18em\optbar{\kern -0.18em \PLambda}{}\xspace}

\newcommand{\decay}[2]{\ensuremath{#1\!\to #2}\xspace}         

\def\to                 {\ensuremath{\rightarrow}\xspace}

\def\CP                {{\ensuremath{C\!P}}\xspace}

\newcommand{\dm}{{\ensuremath{\Delta m}}\xspace}

\def\AT#1     {\ensuremath{A_{\mathrm{T}}^{#1}}\xspace}           

\def\C#1      {\ensuremath{\mathcal{C}_{#1}}\xspace}                       
\def\Cp#1     {\ensuremath{\mathcal{C}_{#1}^{'}}\xspace}                    
\def\Ceff#1   {\ensuremath{\mathcal{C}_{#1}^{\mathrm{(eff)}}}\xspace}        
\def\Cpeff#1  {\ensuremath{\mathcal{C}_{#1}^{'\mathrm{(eff)}}}\xspace}       
\def\Ope#1    {\ensuremath{\mathcal{O}_{#1}}\xspace}                       
\def\Opep#1   {\ensuremath{\mathcal{O}_{#1}^{'}}\xspace}                    

\newcommand{\tev}{\ifthenelse{\boolean{inbibliography}}{\ensuremath{~T\kern -0.05em eV}\xspace}{\ensuremath{\mathrm{\,Te\kern -0.1em V}}}\xspace}
\newcommand{\gev}{\ensuremath{\mathrm{\,Ge\kern -0.1em V}}\xspace}
\newcommand{\mev}{\ensuremath{\mathrm{\,Me\kern -0.1em V}}\xspace}
\newcommand{\kev}{\ensuremath{\mathrm{\,ke\kern -0.1em V}}\xspace}
\newcommand{\ev}{\ensuremath{\mathrm{\,e\kern -0.1em V}}\xspace}
\newcommand{\gevc}{\ensuremath{{\mathrm{\,Ge\kern -0.1em V\!/}c}}\xspace}
\newcommand{\mevc}{\ensuremath{{\mathrm{\,Me\kern -0.1em V\!/}c}}\xspace}
\newcommand{\gevcc}{\ensuremath{{\mathrm{\,Ge\kern -0.1em V\!/}c^2}}\xspace}
\newcommand{\gevgevcccc}{\ensuremath{{\mathrm{\,Ge\kern -0.1em V^2\!/}c^4}}\xspace}
\newcommand{\mevcc}{\ensuremath{{\mathrm{\,Me\kern -0.1em V\!/}c^2}}\xspace}

\def\mum  {\ensuremath{{\,\upmu\rm m}}\xspace}

\def\invfb   {\ensuremath{\mbox{\,fb}^{-1}}\xspace}

\newcommand{\chisq}{\ensuremath{\chi^2}\xspace}

\def\gsim{{~\raise.15em\hbox{$>$}\kern-.85em
          \lower.35em\hbox{$\sim$}~}\xspace}
\def\lsim{{~\raise.15em\hbox{$<$}\kern-.85em
          \lower.35em\hbox{$\sim$}~}\xspace}

\def\sPlot{\mbox{\em sPlot}\xspace}

\def\ptot       {\mbox{$p$}\xspace}
\def\pt         {\mbox{$p_{\rm T}$}\xspace}

\def\evtgen     {\mbox{\textsc{EvtGen}}\xspace}

\def\geant      {\mbox{\textsc{Geant4}}\xspace}

\def\photos     {\mbox{\textsc{Photos}}\xspace}

\def\pythia     {\mbox{\textsc{Pythia}}\xspace}

\def\tell1  {TELL1\xspace}
\def\ukl1   {UKL1\xspace}

\newcommand{\eg}{\mbox{\itshape e.g.}\xspace}

\usepackage{cite} 
\usepackage{mciteplus}

\def\pvalue     {\ensuremath{p}-value\xspace}
\def\pvalues    {\ensuremath{p}-values\xspace}

\def\sWeights{\mbox{\em sWeights}\xspace}

\usepackage{longtable} 
\usepackage[normalem]{ulem}

\begin{document}

\renewcommand{\thefootnote}{\fnsymbol{footnote}}
\setcounter{footnote}{1}

\onecolumn

\begin{titlepage}
\pagenumbering{roman}

\vspace*{-1.5cm}
\centerline{\large EUROPEAN ORGANIZATION FOR NUCLEAR RESEARCH (CERN)}
\vspace*{1.5cm}
\hspace*{-0.5cm}
\begin{tabular*}{\linewidth}{lc@{\extracolsep{\fill}}r}
\ifthenelse{\boolean{pdflatex}}
{\vspace*{-2.7cm}\mbox{\!\!\!\includegraphics[width=.14\textwidth]{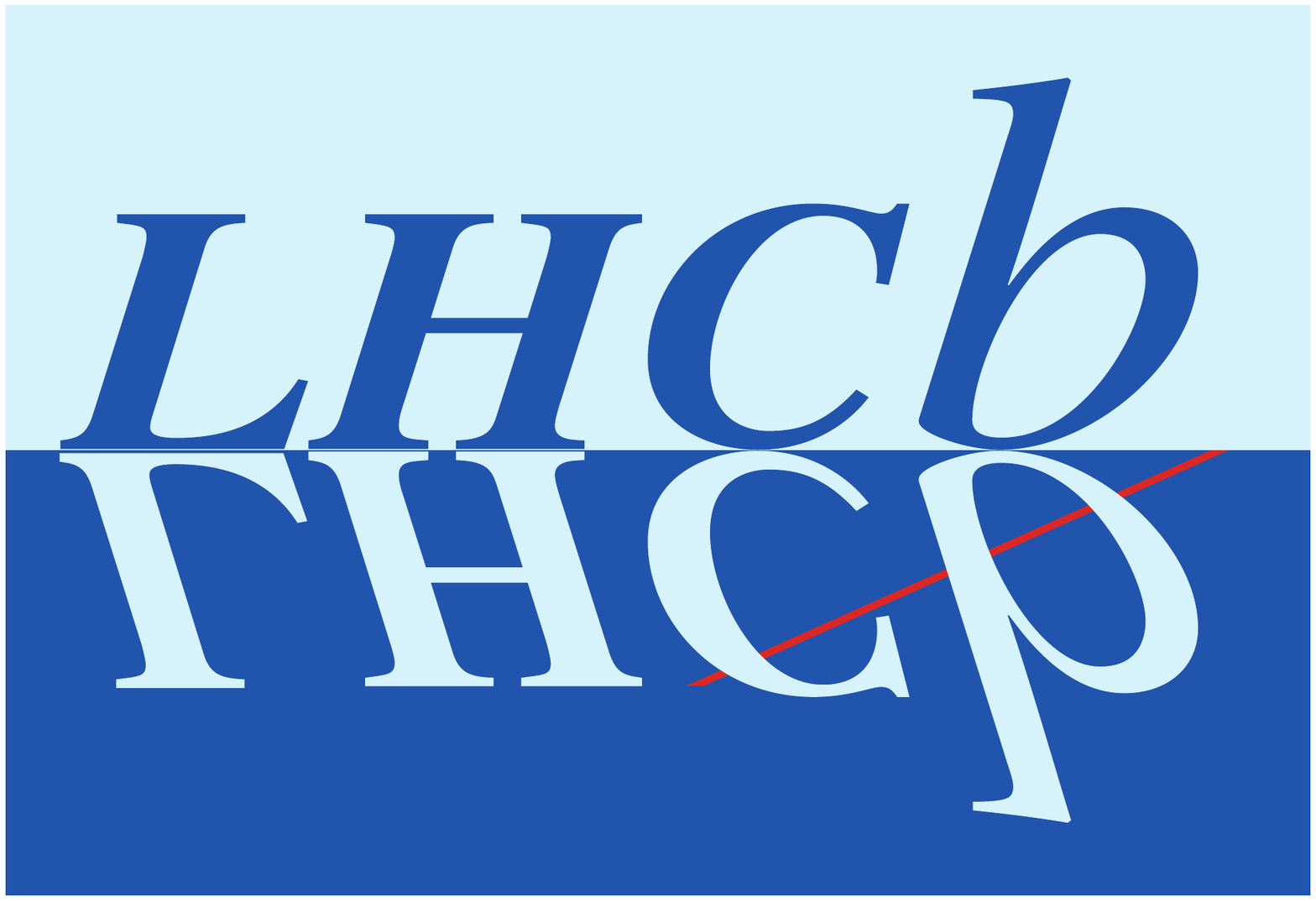}} & &}%
{\vspace*{-1.2cm}\mbox{\!\!\!\includegraphics[width=.12\textwidth]{lhcb-logo.eps}} & &}%
\\
 & & CERN-PH-EP-2014-251 \\  
 & & LHCb-PAPER-2014-054 \\  
 & & 14 October 2014\\ 
\end{tabular*}

\vspace*{4.0cm}

{\bf\boldmath\huge
\begin{center}
Search for \CP violation in \decay{\Dz}{\pim\pip\piz} decays 
with the energy test
\end{center}
}

\vspace*{2.0cm}

\begin{center}
The LHCb collaboration\footnote{Authors are listed at the end of this paper.}
\end{center}

\vspace{\fill}

\begin{abstract}
  \noindent
A search for time-integrated \CP violation in the Cabibbo-suppressed decay 
\decay{\Dz}{\pim\pip\piz} is performed using for the first time 
an unbinned model-independent technique known as the energy test. 
Using proton-proton collision data, corresponding to an integrated luminosity of 
$2.0\invfb$ collected by the \lhcb detector at a centre-of-mass energy of $\sqrt{s} = 8\tev$, 
the world's best sensitivity to \CP violation in this decay is obtained. 
The data are found to be consistent with the hypothesis of 
\CP symmetry with a \pvalue of $(2.6 \pm 0.5)\%$.
\end{abstract}

\vspace*{2.0cm}

\begin{center}
  Submitted to Phys.~Lett.~B
\end{center}

\vspace{\fill}

{\footnotesize 
\centerline{\copyright~CERN on behalf of the \lhcb collaboration, license \href{http://creativecommons.org/licenses/by/4.0/}{CC-BY-4.0}.}}
\vspace*{2mm}

\end{titlepage}


\newpage
\setcounter{page}{2}
\mbox{~}
%
%
%
%

\cleardoublepage

\twocolumn

\renewcommand{\thefootnote}{\arabic{footnote}}
\setcounter{footnote}{0}



\pagestyle{plain} 
\setcounter{page}{1}
\pagenumbering{arabic}


%

\section{Introduction}
\label{sec:Introduction}

The decay \decay{\Dz}{\pim\pip\piz} (charge conjugate decays are implied unless stated otherwise) 
proceeds via a singly Cabibbo-suppressed \decay{\cquark}{\dquark\uquark\dquarkbar} 
transition with a possible admixture from a penguin amplitude.
The interference of these amplitudes may give rise to a violation of the charge-parity symmetry 
(\CP violation), which may be observed as an asymmetry in the total rates, 
or in the distribution of events over the Dalitz plot.
Contributions from particles that are not described in the Standard Model (SM) and participate in the loops 
of the penguin amplitude can enhance the {\it O}($10^{-3}$) \CP violation effects expected within 
the SM~\cite{Grossman:2006jg}. Therefore \CP violation in \decay{\Dz}{\pim\pip\piz} 
decays provides sensitivity to such non-SM physics.

In addition to this direct \CP violation, time-integrated
\CP asymmetry in \decay{\Dz}{\pim\pip\piz} decays can also receive an indirect contribution 
arising from either the \Dz-\Dzb mixing or interference in decays following mixing. 
While direct \CP asymmetry depends on the decay mode, indirect \CP violation is expected 
to be the same for all \CP eigenstates. Recent time-dependent measurements 
of \decay{\Dz}{\pim\pip},~\Km\Kp decays constrain the indirect \CP asymmetry 
to the {\it O}($10^{-3}$) level~\cite{LHCb-PAPER-2013-054}.

The decay \decay{\Dz}{\pim\pip\piz} is dominated by the $\Prho(770)$ resonances, 
with the \Prho meson decaying into a pair of pions, which lead to the final states $\rhoz\piz$, 
$\rhop\pim$, and $\rhom\pip$.
Higher-mass $\rho$ and $f_0$ resonances, as well as $f_2(1270)$ and $\sigma(400)$ particles 
only contribute with fractions at the percent level or less~\cite{Aubert:2007ii}.

Singly Cabibbo-suppressed charm decays have recently received 
a significant attention in the literature (see \eg\ Ref.~\cite{Grossman:2006jg}). 
A particular interest in the decay \decay{\Dz}{\pim\pip\piz} was pointed out in Ref.~\cite{Grossman:2012eb}, 
where the authors derived isospin relations between the different 
$\decay{\D}\Prho\Ppi$ amplitudes and discussed the possibility of identifying contributions 
from non-SM physics using \CP violation measurements.

Previously, the most sensitive search for \CP violation in this decay was performed by 
the \babar collaboration~\cite{Aubert:2008yd}. 
Their result excluded \CP-violating effects larger than a few percent. 
The results presented here are based on a signal sample that is about eight times 
larger and have higher precision.
This is the first \CP violation analysis performed at the \lhcb experiment 
with decays involving \piz mesons.

As \pim\pip\piz is a self-conjugate final state and accessible 
to both \Dz and \Dzb decays, flavour tagging of the \D mesons is performed through 
the measurement of the soft pion ($\pi_s$) charge in the $\decay{\Dstarp}{\Dz\pi_s^+}$ decay.

The method exploited in this Letter, called the energy 
test~\cite{doi:10.1080/00949650410001661440,Aslan2005626}, 
verifies the compatibility of the observed data with \CP symmetry.
It is sensitive to local \CP violation in the Dalitz plot and not to global asymmetries. 
The unbinned technique applied here is used for the first time. 
A visualisation method is also used that allows identification of regions 
of the Dalitz plot in which \CP violation is observed. 
As this model-independent method cannot identify which amplitudes 
contribute to the observed asymmetry, a model-dependent analysis would 
be required in the case of a signal for a non-zero \CP asymmetry.

\section{Detector and reconstruction}
\label{sec:data}

The \lhcb detector~\cite{Alves:2008zz} is a single-arm forward
spectrometer covering the \mbox{pseudorapidity} range $2<\eta <5$,
designed for the study of particles containing \bquark or \cquark
quarks. The detector includes a high-precision tracking system
consisting of a silicon-strip vertex detector surrounding the $pp$
interaction region, a large-area silicon-strip detector located
upstream of a dipole magnet with a bending power of about
$4{\rm\,Tm}$, and three stations of silicon-strip detectors and straw
drift tubes placed downstream of the magnet. 
The combined tracking system provides a measurement of momentum, \ptot,  with
a relative uncertainty that varies from 0.4\% at low momentum to 0.6\% at 100\gevc.
The minimum distance of a track to a primary vertex, the impact parameter (IP), 
is measured with a resolution of $(15+29/\pt)\mum$,
where \pt is the component of \ptot transverse to the beam, in \gevc.
Charged hadrons are identified using two ring-imaging Cherenkov detectors. 
Photon, electron and hadron candidates are identified by a calorimeter system 
consisting of scintillating-pad and preshower detectors, an electromagnetic
calorimeter and a hadronic calorimeter. 
Muons are identified by a system composed of alternating layers of iron 
and multiwire proportional chambers.

The trigger consists of a hardware stage, based on high-\pt signatures  
from the calorimeter and muon systems, followed by a two-level software stage, 
which applies partial event reconstruction. 
The software trigger at its first level requires at least one good quality track associated 
with a particle having high \pt and high $\chi^2_{\rm IP}$, defined as the difference in $\chi^2$ 
of the primary $pp$ interaction vertex (PV) reconstructed with and without this particle. 

A dedicated second-level trigger partially reconstructs \decay{\Dz}{\pim\pip\piz} candidates coming from 
$\decay{\Dstarp}{\Dz\pi_s^+}$ decays using only information from the charged particles. 
These requirements ensure the suppression of combinatorial background without 
distorting the acceptance in the decay phase space.
The \pim\pip pair is combined with a pion to form a \Dstarp candidate, which is accepted if it
has \pt greater than $2.5\gevc$ and a difference of invariant masses 
$m(\pim\pip\pi_s^+)-m(\pim\pip)<285\mevcc$. All the charged-particle tracks used must have good quality, 
$\pt>0.3\gevc$ and $p>3.0\gevc$, while the charged pions from the \Dz decays must also 
have high $\chi^2_{\rm IP}$. The \pim\pip combination is required to form a good quality 
secondary vertex significantly displaced from the PV, while the soft pion must originate from the PV. 

The inclusive \Dstarp trigger was introduced at the start of 2012.
The data collected by the LHCb experiment in 2012 corresponds to an integrated luminosity 
of $2\invfb$ of $pp$ collisions collected at a centre-of-mass energy of $8\tev$. 
The magnetic field polarity is reversed regularly during the data taking 
with approximately half the data collected at each polarity to reduce the overall effect of any charge-dependent detection and reconstruction efficiency.

In this analysis, the two \piz categories reconstructed in LHCb are exploited~\cite{LHCb-DP-2014-002}.  
These are pions for which both final state photons are reconstructed separately (resolved pions), 
as well as pions that have higher momentum (typically $\pt>2\gevc$)
and thus a smaller opening angle of the two photons (merged pions). 
These \piz mesons are detected in the calorimeter as one merged cluster 
which is further split into two subclusters based on the expected shape of the photon shower.
The merged pions make up about 30\% of the reconstructed \piz mesons.
Among the resolved \piz mesons there are also candidates made of photons which, 
after interacting with detector material, have converted into an $e^+e^-$ pair. 
The two \piz samples provide coverage of complementary regions 
of the \decay{\Dz}{\pim\pip\piz} Dalitz plot and thus the use of both contributes significantly 
to the sensitivity of the analysis.

\section{Event selection}
\label{sec:selection}

The offline selection is split into a pre-selection, which follows the trigger selection, 
and a selection based on a boosted decision tree (BDT)~\cite{Breiman,AdaBoost}. 
All the \Dstarp candidates are required to pass both levels of the software trigger.
In addition, the pre-selection requires more stringent kinematic criteria than those applied in the trigger;
in particular $\pt>0.5\gevc$ is required for all the \Dz decay products to reduce the combinatorial background.
For resolved \piz mesons, the di-photon invariant mass has to be within $15\mevcc$ of the known \piz mass, 
this corresponds to about three times the $m(\gamma \gamma)$ resolution.
The invariant mass of merged photons, due to its lower resolution, is required 
to be within the range of $75-195~\mevcc$. The purity of the merged \piz sample is nevertheless 
significantly higher with respect to the resolved \piz, as it benefits from large 
transverse energy and much lower combinatorial background.

Cross-feed from \decay{\Dz}{\Km\pip\piz} decays, with a kaon misidentified as a pion, 
is reduced with requirements on the \pipm particle identification 
based on the ring-imaging Cherenkov detectors. 

The \Dz candidates satisfying the above criteria and having invariant mass 
$m(\pim\pip\piz)$ within $40 (60) \mevcc$ of the known \Dz mass are accepted 
in the resolved (merged) sample; this range corresponds to approximately 
four times the $m(\pim\pip\piz)$ resolution.
The \Dstarp candidates, formed with the \Dz and $\pi_s^+$ candidate,
have their entire decay chain refitted requiring that the \Dstarp candidate originates 
from the corresponding PV, and \piz and \Dz candidates have their nominal masses. 
This improves the \Dstarp mass resolution and the resolution 
in the \decay{\Dz}{\pim\pip\piz} Dalitz plot, while a requirement put 
on the fit quality efficiently suppresses the background. 
This requirement also suppresses the contribution from the \Dstarp mesons originating 
from long-lived \bquark-hadrons. The remainder of this component is not affected by \CP asymmetries 
in \bquark-hadrons since the flavour tag is obtained from the \Dstarp meson.

\begin{figure}[tbh]
  \centering
  \includegraphics[height=0.38\textwidth]{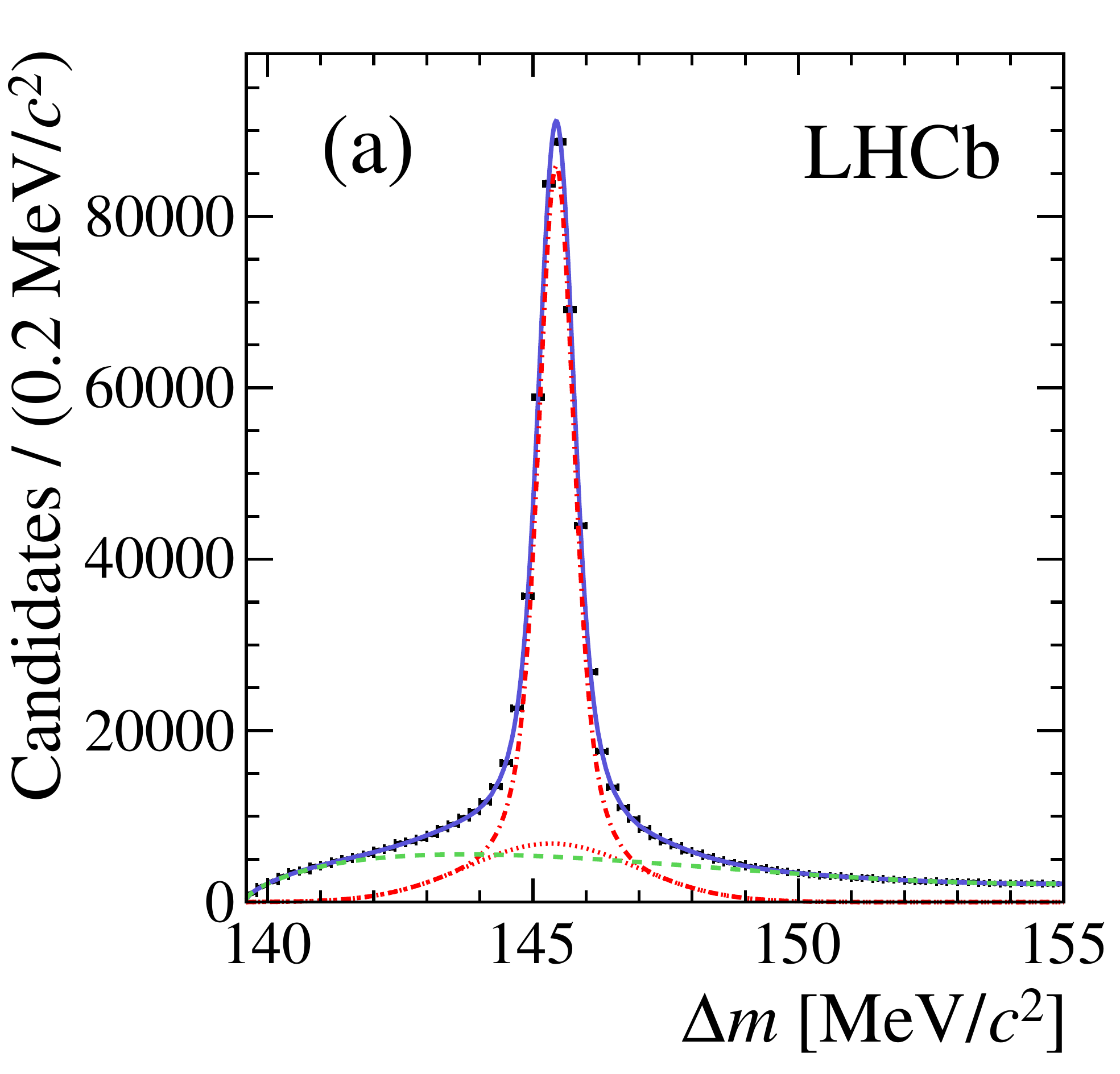}
  \includegraphics[height=0.38\textwidth]{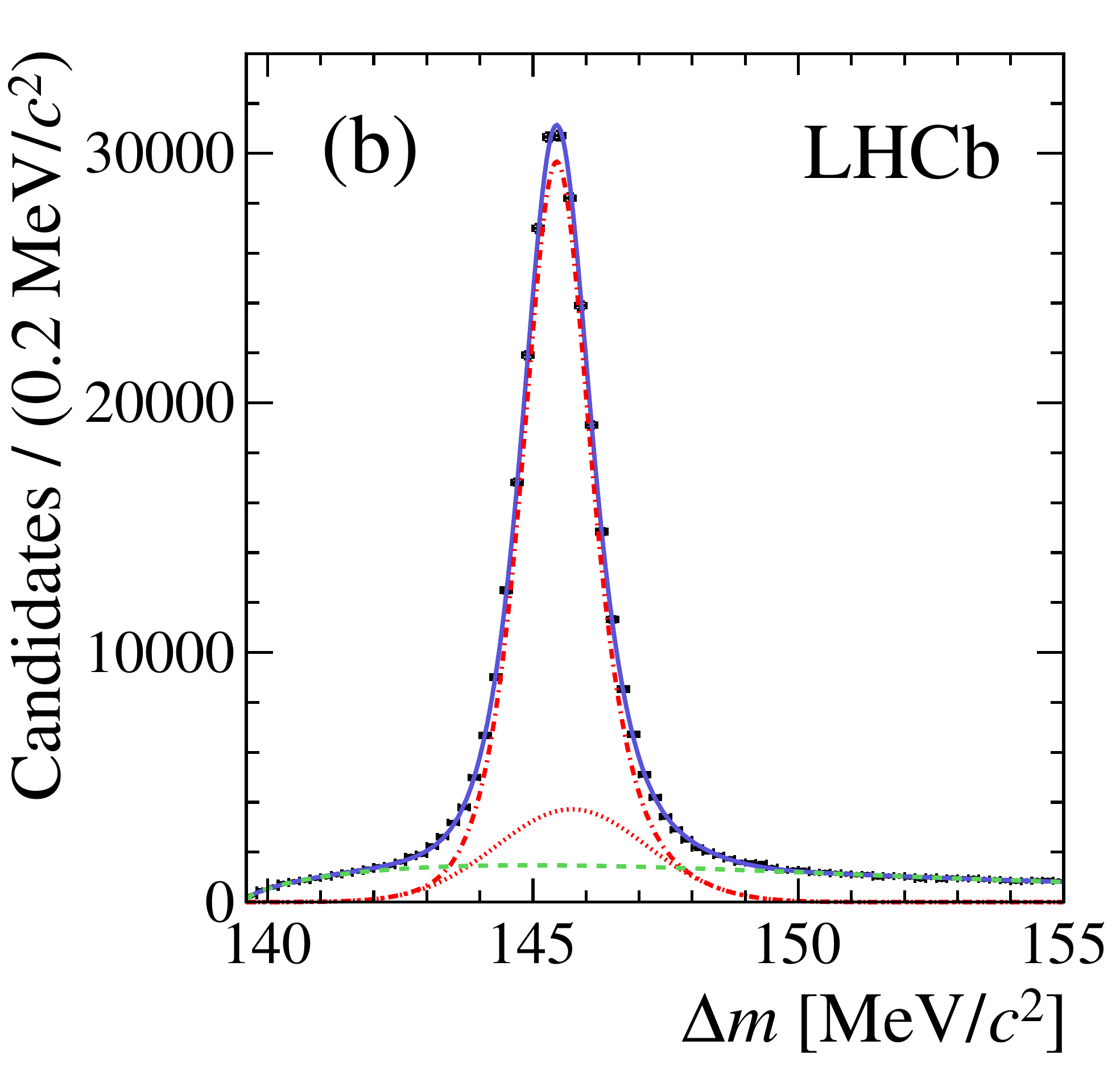}
  \caption{Distribution of $\Delta{}m$ with fit overlaid for the selected data set with (a) resolved  
  and (b) merged \piz candidates. The lines show the fit results for total signal (dot-dashed red), 
  widest Gaussian signal component (dotted red), background (dashed green), and total (solid blue).
  \label{fig:Data}}
\end{figure}

This preliminary selection is followed by a multivariate analysis based on a BDT. 
Signal and background samples used to train the BDT are obtained by applying 
the \sPlot technique~\cite{Pivk:2004ty} to a quarter of the real data. The \sWeights 
for signal and background separation are determined from a fit to the distribution 
of the mass difference, $\dm \equiv m(\pim\pip\piz \ \pi_s^+) - m(\pim\pip\piz)$, 
separately for the resolved and merged samples. 
The BDT uses the variables related to the kinematic and topological properties
of the signal decays, as well as the \piz quality. 
It is trained separately for the resolved and merged data categories. 
The most discriminating variables in the resolved sample 
are $\pt(\pi_s^+)$, $\pt(\Dz)$ and $\pt(\piz)$, while in the merged sample 
these are $\pt(\pi_s^+)$, $\pt(\Dz)$ and the \Dz $\chisq_{\rm IP}$. The optimal 
value of the BDT discriminant is determined by estimating the \Dstarp signal 
significance for various requirements on the BDT output. 
It retains approximately 75\% (90\%) of the resolved (merged) 
signal events while removing 90\% (55\%) of the background.
Figure~\ref{fig:Data} shows the $\Delta{}m$ distributions for the selected 
data set for events with resolved and merged \piz candidates. 
The signal shapes, fitted in Fig.~\ref{fig:Data} with a sum of three Gaussian functions, 
significantly differ between both samples reflecting the different \piz momentum resolutions.  
The lower momentum resolution of the merged \piz mesons relative to the resolved \piz mesons makes 
the core part of the merged signal distribution wider, while the low-\pt \ \piz mesons 
contributing to the resolved signal enlarge its tail component.
The background shape is fitted using a second-order polynomial multiplied by $\sqrt{1-m_{\pip}/\dm}$.

\begin{figure*}[tb]
  \centering
  \includegraphics[width=0.34\textwidth]{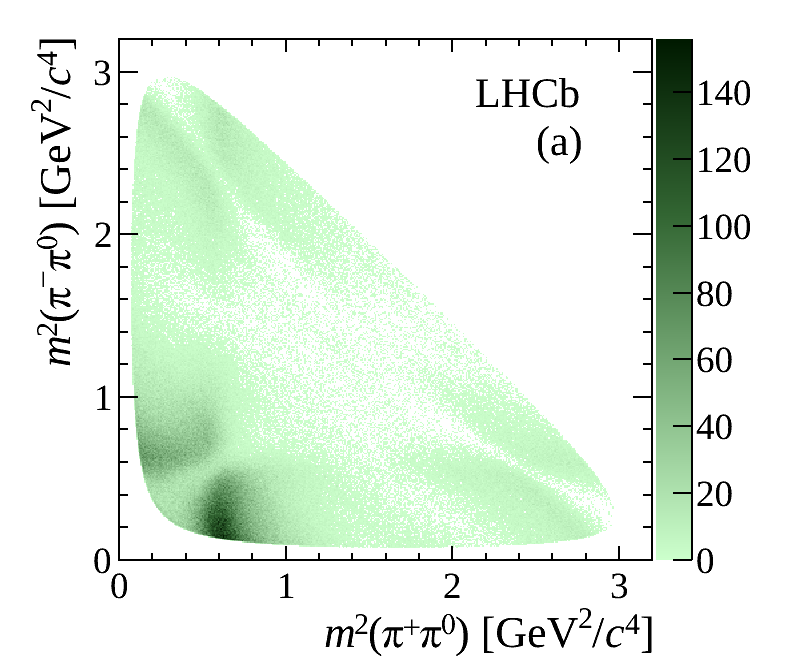}
  \hspace{-0.03\textwidth}
  \includegraphics[width=0.34\textwidth]{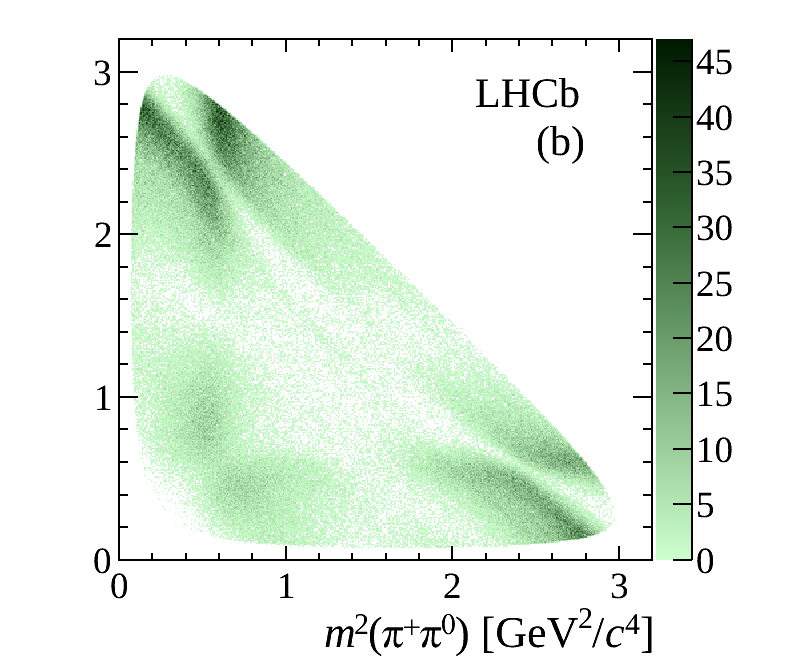}
  \hspace{-0.03\textwidth}
  \includegraphics[width=0.34\textwidth]{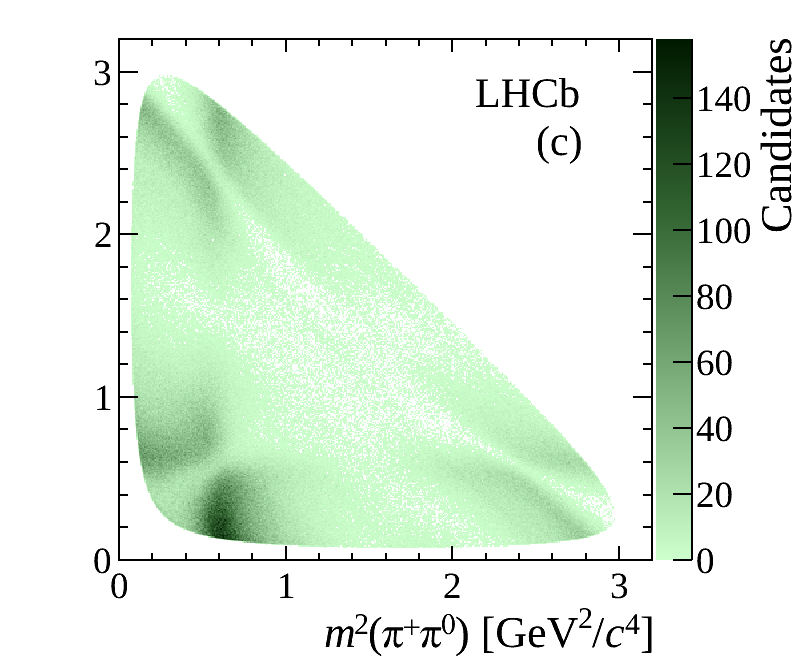}
  \caption{Dalitz plot of the (a) resolved, (b) merged and (c) combined 
  \decay{\Dz}{\pim\pip\piz} data sample. Enhanced event densities 
  in the phase-space corners originate from the $\rho(770)$ resonances.
  \label{fig:Data1}}
\end{figure*}

The final signal sample is selected requiring $|\dm -145.4| <1.8 \mevcc$, 
which corresponds 
to roughly four times the effective $\Delta{}m$ resolution. 
The effective resolution is similar for both resolved and merged \piz samples when averaging the narrow and broad components of the peak.
This gives $416 \times 10^3$ resolved and $247 \times 10^3$ 
merged signal candidates with a purity of $82\%$ and $91\%$, respectively.
The Dalitz plot of the final signal sample is shown in Fig.~\ref{fig:Data1}.
The smaller number of candidates in the low $m^2(\pip \pim)$ region compared with 
the high $m^2(\pip \pim)$ region is due to acceptance effects related to 
the \piz reconstruction as discussed in Sec.~\ref{sec:sensitivity}.

\section{Energy test method}
\label{sec:method}

Model-independent searches for local \CP violation are typically carried out using a binned \chisq  
approach to compare the relative density in the Dalitz plot of 
a decay and its \CP-conjugate sample (see for example \cite{Aubert:2008yd,Bediaga:2009tr}). 
A model-independent unbinned statistical method called the energy test was introduced in Refs.~\cite{doi:10.1080/00949650410001661440,Aslan2005626}. 
Reference~\cite{Williams:2011cd} suggests applying this method to Dalitz plot analyses 
and demonstrates the potential to obtain improved sensitivity to \CP violation over 
the standard binned approach. This Letter describes the first application of this technique 
to experimental data.

In this method a test statistic, $T$, is used to compare average distances in phase space, 
based on a metric function, $\psi_{ij}$, of pairs of events $ij$ 
belonging to two samples of opposite flavour.
It is defined as
\begin{equation}
T = \sum_{i,j>i}^{n}\dfrac{\psi_{ij}}{n(n-1)}
 + \sum_{i,j>i}^{\overline{n}}\dfrac{\psi_{ij}}{\overline{n}(\overline{n}-1)}
 - \sum_{i,j}^{n,\overline{n}}\dfrac{\psi_{ij}}{n\overline{n}} ,
\label{eqn:T}
\end{equation}
where the first and second terms correspond to a metric-weighted average distance 
of events within $n$ events of one flavour and $\overline{n}$ events of the opposite flavour, respectively. 
The third term measures the weighted average distance of events in one flavour sample to events 
of the opposite flavour sample. The normalisation factors in the denominator 
remove the impact of global asymmetries.
If the distributions of events in both flavour samples are identical, $T$ will fluctuate around a value close to zero. 

The metric function should be falling with increasing distance $d_{ij}$ between events $i$ and $j$, 
in order to increase the sensitivity to local asymmetries. A Gaussian metric is chosen, defined as 
$\psi_{ij}\equiv\psi(d_{ij})=e^{-d_{ij}^2/2\sigma^2}$ with a tunable parameter $\sigma$, which
describes the effective radius in phase space within which 
a local asymmetry is measured. Thus, this parameter should be larger than the resolution of $d_{ij}$ 
and small enough not to dilute locally varying asymmetries.

The distance between two points in phase space is usually measured as the distance in the Dalitz plot.
However, this distance depends on the choice of the axes of the Dalitz plot.
This dependence is removed by using all three invariant masses to determine 
the distance, $d_{ij}$, calculated as the length of the displacement vector
$\Delta\vec{x}_{ij}=(m_{12}^{2,j}-m_{12}^{2,i},m_{23}^{2,j}-m_{23}^{2,i},m_{13}^{2,j}-m_{13}^{2,i})$, 
where the $1,2,3$ subscripts indicate the final-state particles. 
Using all three invariant masses does not add information, 
but it avoids an arbitrary choice that could impact the sensitivity of the method to different 
\CP violation scenarios.

In the case of \CP violation, the average distances entering in the third term 
of Eq.~\ref{eqn:T} are larger, which, because of the characteristics of the metric function, 
leads to a reduced magnitude of this term. Therefore larger \CP asymmetries lead to larger values of $T$.
This is translated into a \pvalue under the hypothesis of \CP symmetry by comparing 
the nominal $T$ value observed in data to a distribution of $T$ values obtained from permutation 
samples, where the flavour of each candidate is randomly reassigned to simulate samples without \CP violation.
The \pvalue for the no \CP violation hypothesis is obtained as the fraction of permutation $T$ values greater 
than the nominal $T$ value. 

\begin{figure*}[bt]
\centering
    \includegraphics[width=0.49\textwidth]{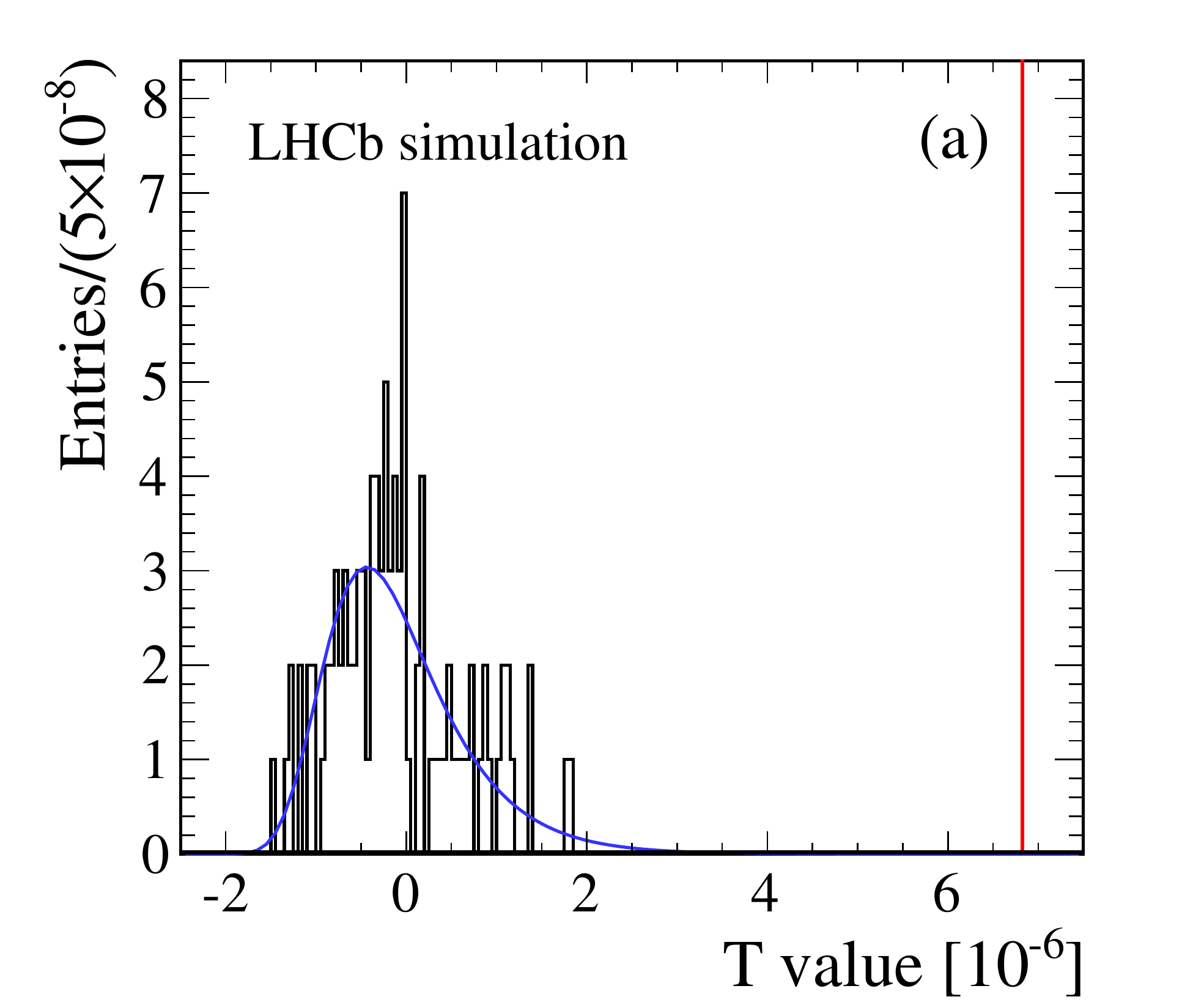}
    \includegraphics[width=0.49\textwidth]{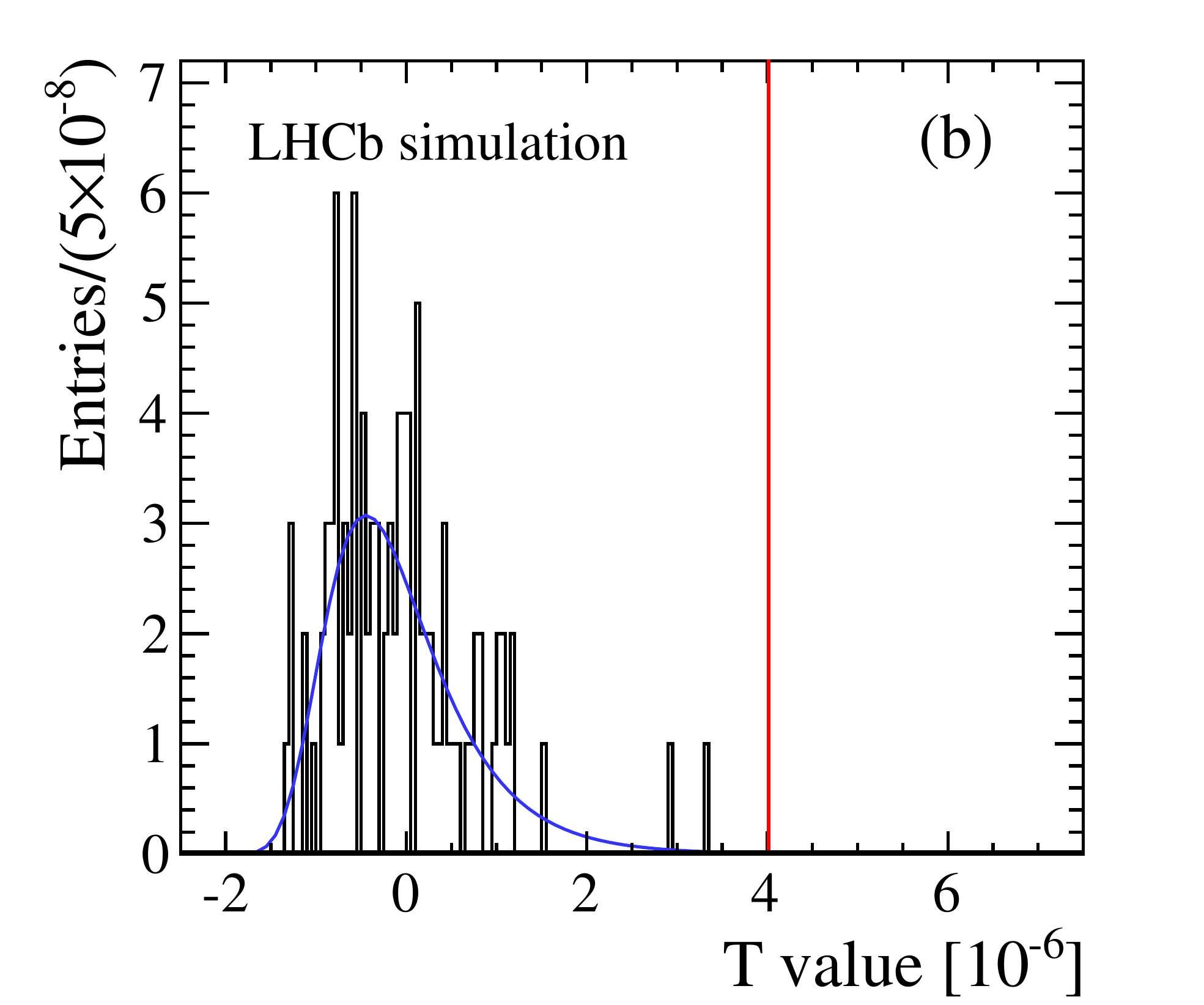}
  \caption{
    \small 
    Distribution of permutation $T$ values fitted with a GEV function for the simulated sample and showing 
    the nominal $T$ value as a vertical line for (a) $2\%$ \CP violation in the amplitude and (b) 
    $1^\circ$ phase \CP violation of the $\rho^+$ resonance.}
  \label{fig:method_visualisation_1}
\end{figure*}

A statistical uncertainty of the \pvalue is obtained as a binomial standard deviation.
If large \CP violation is observed, the observed $T$ value is likely to lie outside 
the range of permutation $T$ values. In this case the permutation $T$ distribution can be fitted 
with a generalised extreme value (GEV) function, as demonstrated in 
Refs.~\cite{doi:10.1080/00949650410001661440,Aslan2005626} and verified in large simulation samples 
for this analysis.
The GEV function is defined as
\begin{eqnarray}
f(T;\mu,\delta,\xi) = N \left[1+\xi\left(\frac{T-\mu}{\delta}\right)\right]^{(-1/\xi)-1}&&\nonumber\\
\times\exp\left\{-\left[1+\xi\left(\frac{T-\mu}{\delta}\right)\right]^{-1/\xi}\right\},&&
\end{eqnarray}
with normalisation $N$, location parameter $\mu$, scale parameter $\delta$, and shape parameter $\xi$.
This function is set to zero for $T>\mu-\delta/\xi$ for $\xi>0$, and for $T<\mu-\delta/\xi$ for $\xi<0$.
Figure~\ref{fig:method_visualisation_1} shows an example $T$ value distribution with a GEV function 
fit for a simulated data set including \CP violation (see Sec.~\ref{sec:sensitivity}). 

\begin{figure*}[tb]
\centering
    \includegraphics[width=0.49\textwidth]{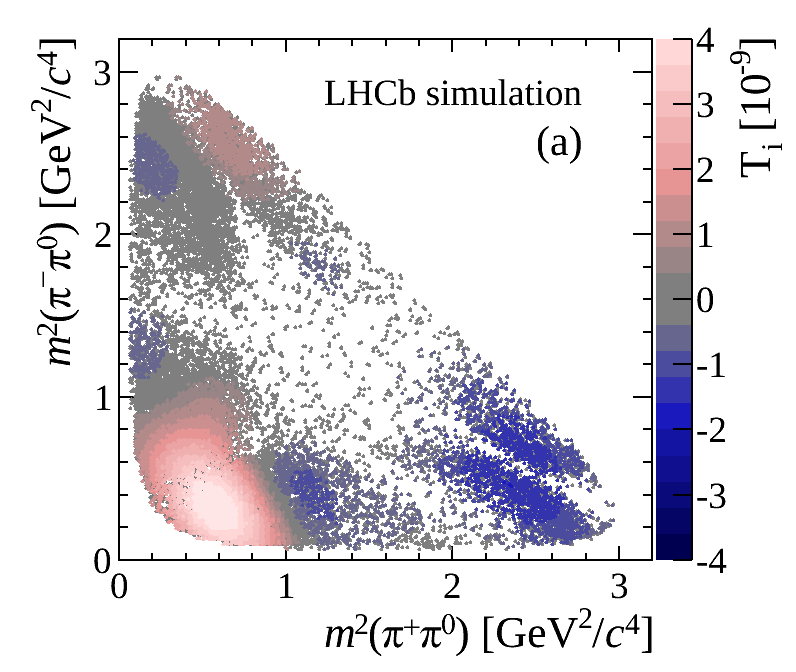}
    \includegraphics[width=0.49\textwidth]{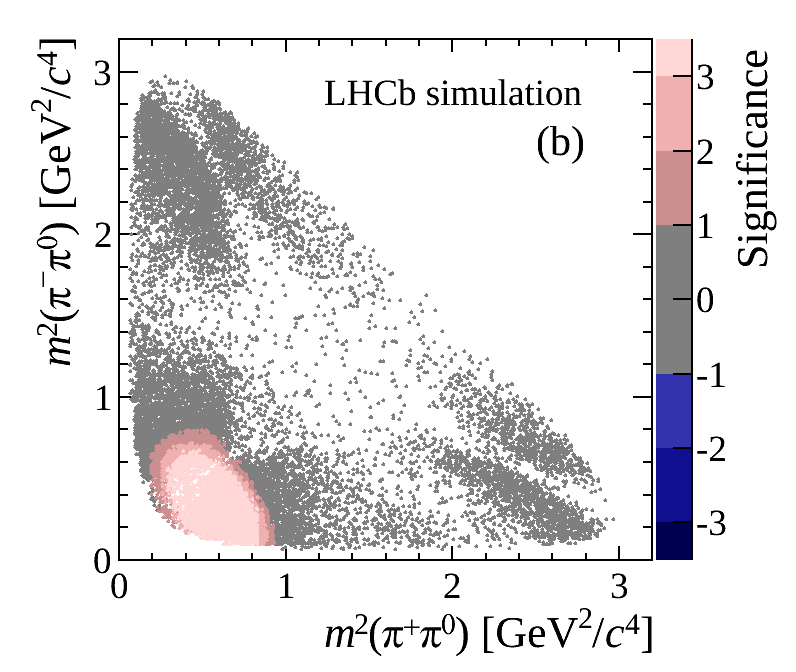}\\
    \includegraphics[width=0.49\textwidth]{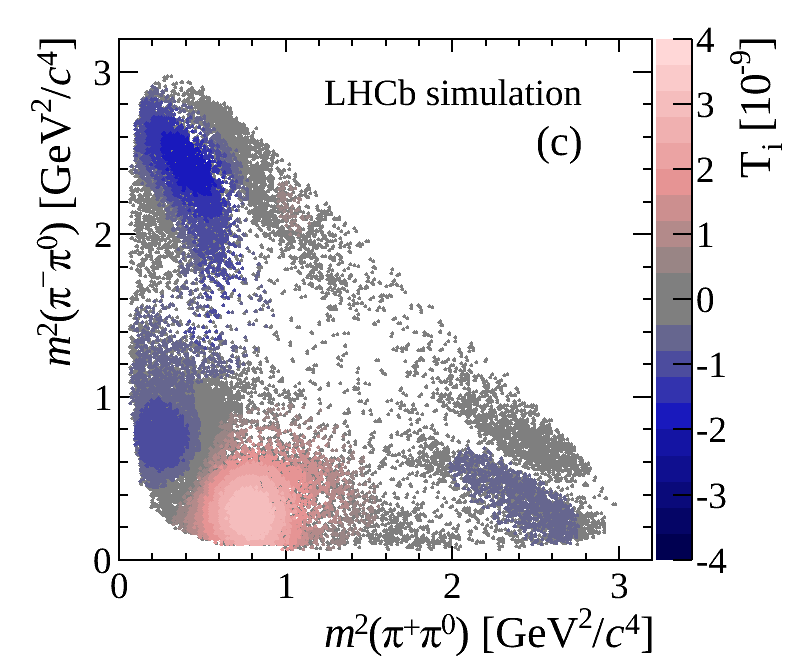}
    \includegraphics[width=0.49\textwidth]{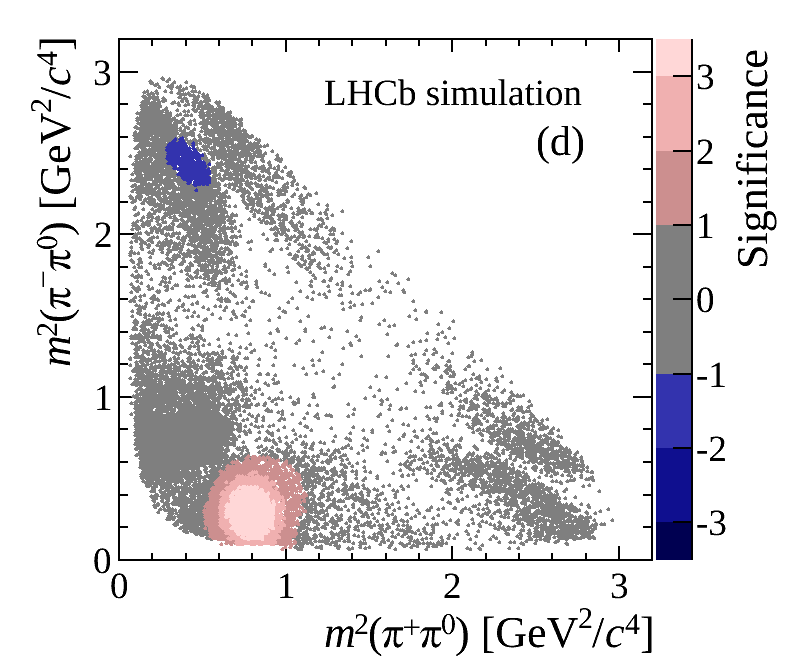}\\
  \caption{
    \small 
    (a,c) $T_i$ value distributions, 
    and (b,d) local asymmetry significances for (top) $2\%$ \CP violation in the amplitude and (bottom) 
    $1^\circ$ phase \CP violation of the $\rho^+$ resonance.}
  \label{fig:method_visualisation}
\end{figure*}

The \pvalue from the fitted $T$ distribution can be calculated as the fraction of the integral 
of the function above the nominal $T$ value.
The uncertainty on the \pvalue is obtained by randomly resampling the fit parameters 
within their uncertainties, taking into account their correlations, and by extracting a \pvalue for each of 
these generated $T$ distributions. The spread of the resulting \pvalue distribution is used to set 68\% confidence uncertainties. A 90\% confidence upper limit is quoted where no significantly non-zero \pvalue can be obtained 
from the fit.

The number of available permutations is constrained by the available computing time.
The default \pvalue extraction uses the counting method as long as at least three permutation $T$ values are found to be 
larger than the observed $T$ value. Beyond that, the \pvalue is extracted by integrating the fitted GEV function.
The \pvalues presented here are based on 1000 permutations for the default data results and on 100 permutations 
for the sensitivity studies.

A visualisation of regions of significant asymmetry is obtained by assigning an asymmetry significance 
to each event. The contributions of a single event of one flavour, $T_i$, and a single event of 
the opposite flavour, $\overline{T}_i$, to the total $T$ value are given by
\begin{eqnarray}
\label{eq:Ti}
T_{i} & = & \dfrac{1}{2n\left( n-1\right) }\sum_{j\neq i}^{n}\psi_{ij}-\dfrac{1}{2n\overline{n} }\sum_{j}^{\overline{n}}\psi_{ij},\nonumber\\
\overline{T}_i & = & \dfrac{1}{2\overline{n}\left( \overline{n}-1\right) }\sum_{j\neq i}^{\overline{n}}\psi_{ij}-\dfrac{1}{2n\overline{n} }\sum_{j}^{n}\psi_{ij}.
\end{eqnarray}

Example $T_i$ distributions for the simulated \CP violating data sets are shown in 
Fig.~\ref{fig:method_visualisation} (a,c); events contributing with $T_i$ of 
the largest magnitude point to \CP violation regions. However, the \CP asymmetry arising 
from the $\rho^+$ amplitude difference (Fig.~\ref{fig:method_visualisation} (a)) 
produces a global asymmetry, $n>\overline{n}$. Through the normalization factors 
in Eq.~\ref{eq:Ti}, this leads to negative $T_i$ regions for approximately $m^2(\pip\piz)>2$~\gevgevcccc, 
where the numbers of \Dz and \Dzb mesons are equal.

Having obtained the $T_i$ and $\overline{T}_i$ values for all events, a permutation method is also used here to 
define the level of significance. The distributions of the smallest negative
and largest positive $T_i$ values of each permutation, $T_{i}^{\rm min}$ and $T_{i}^{\rm max}$, 
are used to assign significances of negative and positive asymmetries, respectively. 
Positive (negative) local asymmetry significances are $T_i$ values greater (smaller) than the fraction of the 
$T_{i}^{\rm max}$ ($T_{i}^{\rm min}$) distribution that corresponds to the significance level. The same procedure  
is applied to the $\overline{T}_i$ distribution, leading to a Dalitz plot with an inverted asymmetry pattern.

The asymmetry significances for each simulated event are plotted on a Dalitz plot 
(see Fig.~\ref{fig:method_visualisation} (b,d)). 
If an amplitude difference exists between \CP-conjugate states of a resonance, 
the region of significant asymmetry appears as a band around the mass of the resonance on the plot.
If a phase difference is present instead, regions of positive and negative 
asymmetry appear around the resonance on the plot, indicating the phase shift.

The practical limitation of this method is that the number of mathematical operations scales quadratically 
with the sample size. Furthermore, a significant number of permutations is required to get a sufficient 
precision on the \pvalue. In this analysis the method is implemented using parallelisation on graphics 
processing units (GPUs)~\cite{Thrust}. 

\section{Sensitivity studies}
\label{sec:sensitivity}

The interpretation of the results requires a study 
of the sensitivity of the present data sample to different types of \CP violation.
The sensitivity is examined based on simplified Monte Carlo samples generated according to the model 
described in Ref.~\cite{Aubert:2007ii} using the generator package Laura++~\cite{Laura}.

\begin{figure}[t!]
  \centering
  \includegraphics[width=0.49\textwidth]{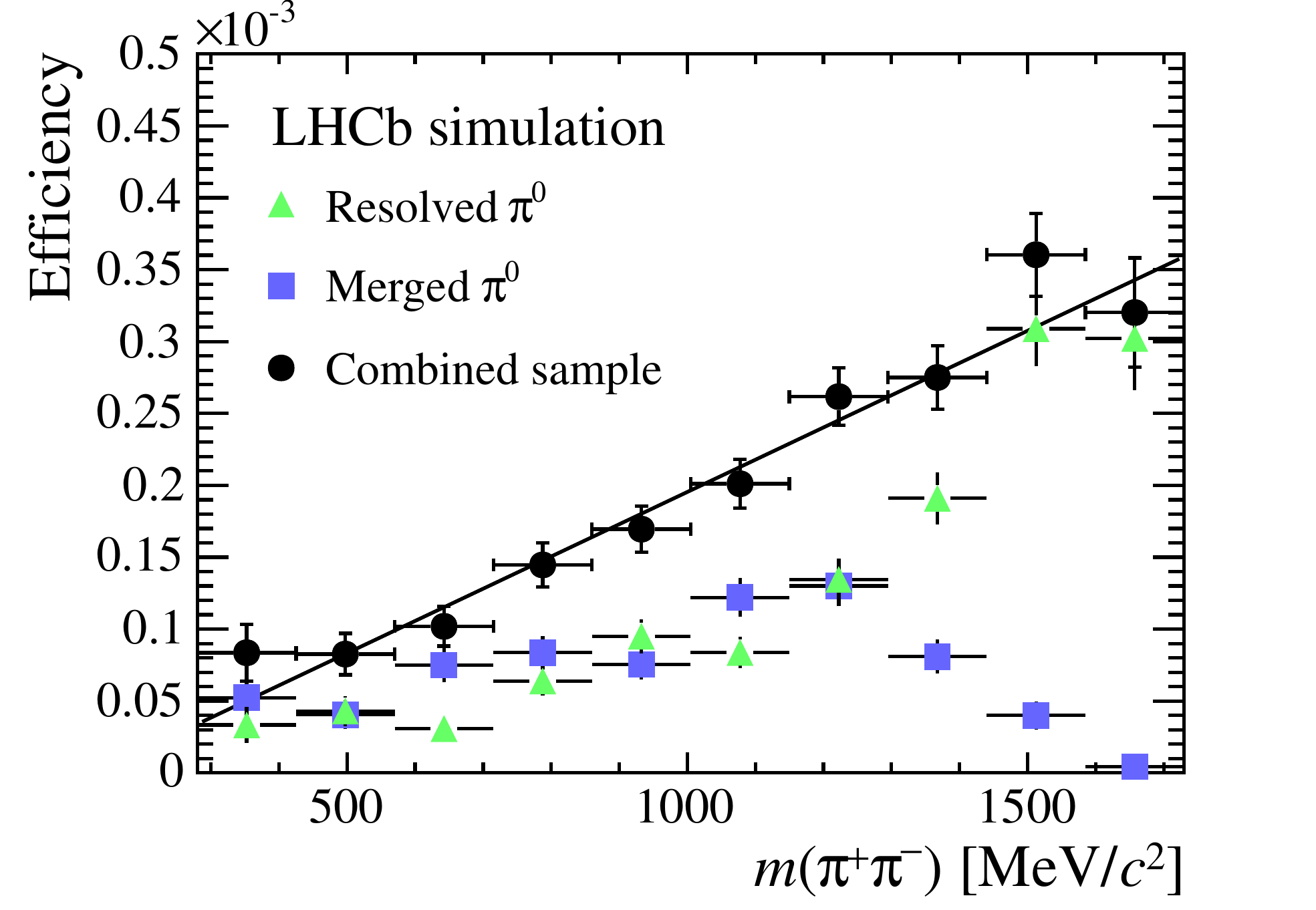}
  \caption{The selection efficiency as a function of $m(\pip\pim)$. 
  The efficiency for the combined sample is fitted with a straight line.
  \label{fig:Efficiency}}
\end{figure}

The selection efficiency has to be taken into account in these studies as it varies 
strongly across phase space. This efficiency is measured using a sample of events 
based on the full \lhcb detector simulation. In the simulation, $pp$ collisions are 
generated using \pythia~\cite{Sjostrand:2006za,*Sjostrand:2007gs} with a specific \lhcb
configuration~\cite{LHCb-PROC-2010-056}. Decays of hadronic particles are described by 
\evtgen~\cite{Lange:2001uf}, in which final-state radiation is generated using 
\photos~\cite{Golonka:2005pn}. The interaction of the generated particles with the detector 
and its response are implemented using the \geant toolkit~\cite{Allison:2006ve, *Agostinelli:2002hh} 
as described in Ref.~\cite{LHCb-PROC-2011-006}.
The efficiency is shown to vary as a function of the $\pip\pim$ invariant mass, while it is found 
to be constant with respect to other variables. The resulting efficiency curves for merged and resolved 
\piz mesons as well as for the combined sample are shown in Fig.~\ref{fig:Efficiency}. 
The small $m(\pip\pim)$ range corresponds to \piz candidates with high momentum 
and is primarily covered by the merged \piz sample. The overall efficiency in this region is low 
as the integrated merged \piz identification efficiency is lower than for the resolved sample and decreases 
after its turn-on as the \piz momentum increases. This affects in particular the $\rho^0$ resonance, as it lies entirely 
within the low acceptance region.

For further studies, the efficiency, based on the fitted curve, is then applied to simplified 
Monte Carlo data sets by randomly discarding events based on the candidate's position in phase space. 
Background events are simulated by resampling phase-space distributions extracted from $\Delta{}m$ 
sideband regions. Inclusion of the background does not significantly reduce 
the sensitivity to the \CP violation scenarios discussed below. 
This is due to the low level of background and due to it being \CP symmetric within 
the present sensitivity.

\begin{table}[b!]
\caption{Overview of sensitivities to various \CP violation scenarios. 
$\Delta A$ and $\Delta \phi$ denote, respectively, 
change in amplitude and phase of the resonance $R$.}
\centering
\begin{tabular}{lll}
\hline \hline 
$R$ ($\Delta A$, $\Delta \phi$) & \pvalue (fit) & Upper limit \\ 
\hline  
$\rho^0$ $(4\%$, $0^\circ)$ & $3.3^{+1.1}_{-3.3}\times10^{-4}$ & $4.6\times10^{-4}$  \\ 
$\rho^0$ $(0\%$, $3^\circ)$ & $1.5^{+1.7}_{-1.4}\times10^{-3}$ & $3.8\times10^{-3}$  \\ 
$\rho^+$ $(2\%$, $0^\circ)$ & $5.0^{+8.8}_{-3.8}\times10^{-6}$ & $1.8\times10^{-5}$  \\ 
$\rho^+$ $(0\%$, $1^\circ)$ & $6.3^{+5.5}_{-3.3}\times10^{-4}$ & $1.4\times10^{-3}$  \\ 
$\rho^-$ $(2\%$, $0^\circ)$ & $2.0^{+1.3}_{-0.9}\times10^{-3}$ & $3.9\times10^{-3}$  \\ 
$\rho^-$ $(0\%$, $1.5^\circ)$ & $8.9^{+22}_{-6.7}\times10^{-7}$ & $4.2\times10^{-6}$  \\ 
\hline \hline  
\end{tabular}
\label{tab:sensitivity_overview}
\end{table}
 
Various \CP asymmetries are introduced by modifying, for a chosen \Dz flavour, 
either the amplitude or the phase of one of the three intermediate $\rho$ resonances 
dominating the $\pim\pip\piz$ phase space. The resulting sensitivities are shown in 
Table~\ref{tab:sensitivity_overview}. The \pvalues, including their statistical uncertainties, 
are obtained from fits of GEV functions 
to the $T$ value distributions and $90\%$ confidence limits are given in addition. 

The sensitivity is comparable to that of the \babar analysis~\cite{Aubert:2008yd} for 
the $\rho^0$ resonance and significantly better for the $\rho^+$ 
and $\rho^-$ resonances. This is expected due to the variation of the selection efficiency across 
phase space, which disfavours the $\rho^0$ region.

The sensitivity of the method also depends on the choice of the metric parameter $\sigma$.
Studies indicate good stability of the measured sensitivity
for values of $\sigma$ between $0.2$ and $0.5\gevgevcccc$, which 
are well above the resolution of the $d_{ij}$ and small compared to the size of the phase space.
The value $\sigma=0.3\gevgevcccc$ yields the best sensitivity to some of the \CP violation 
scenarios studied and was chosen, prior to the data unblinding, as the default value. 
The optimal $\sigma$ value may vary with different \CP violation scenarios. 
Hence the final results are also quoted for several values of $\sigma$.

The standard binned method~\cite{Bediaga:2009tr} is also applied to the 
simulated data sets. This study shows that the energy test provides results 
compatible with, and equally or more precise than the binned method. 

There are two main sources of asymmetry that may degrade or bias the results.
One is an asymmetry that may arise from background events and the other is due to particle detection 
asymmetries that could vary across phase space. 

Background asymmetries are tested by applying the energy test to events in the upper $\Delta{}m$ sideband, 
$\Delta{}m>150$~\mevcc. No significant asymmetry is found. 
In addition, simplified simulation data sets are produced by generating signal candidates 
without \CP violation and background candidates according to background distributions in data, separately 
for \Dz and \Dzb candidates and thus allowing for a background-induced asymmetry. These samples show
a distribution of \pvalues consistent with the absence of any asymmetry.
Further tests using a binned approach~\cite{Bediaga:2009tr} confirm this conclusion.
These are carried out on the $\Delta{}m$ sideband data sample as well as on background samples 
obtained using the \sPlot technique based on the $\Delta{}m$ fits in Fig.~\ref{fig:Data}.
Both approaches show no indication of a background asymmetry.
As the background present in the signal region is found to be \CP symmetric,
it is simply included in the $T$ value calculation discussed in Sec.~\ref{sec:method}.

Local asymmetries are expected to arise, at a level below the current sensitivity, due to 
the momentum dependence of $\pip/\pim$ detection asymmetries in combination with the different kinematic 
distributions of \pip and \pim in certain regions of phase space.

These effects are tested using the Cabibbo-favoured decay \decay{\Dz}{\Km\pip\piz} as a control mode.
This channel is affected by kaon detection asymmetries, which are known to be larger than pion detection 
asymmetries and thus should serve as a conservative test. The data sample is split into eight subsets, 
each of which contains approximately the same amount of data as the signal sample.
The energy test yields \pvalues between $3\%$ and $74\%$, which is consistent with the assumption 
that detection asymmetries are below the current level of sensitivity.
A further test is conducted by splitting the control mode data sample by the polarity of the spectrometer 
dipole magnet, which yields two large approximately equal-sized samples. 
The resulting \pvalues of $8\%$ and $15\%$ show no evidence of sizable biases due to detector asymmetries.

\section{Results and conclusions}
\label{sec:results}

\begin{figure}[htb]
  \centering
  \includegraphics[width=0.49\textwidth]{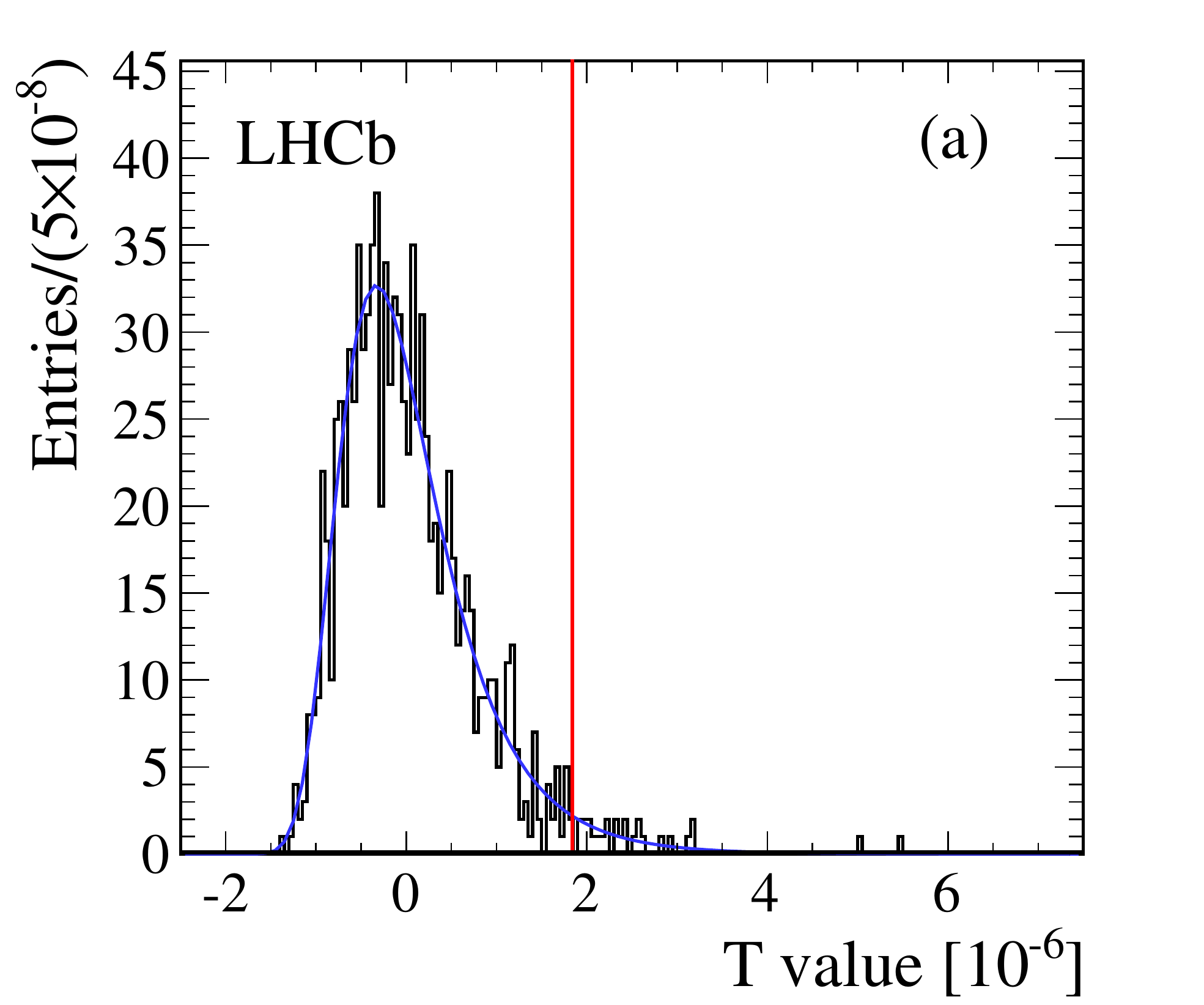}
  \includegraphics[width=0.49\textwidth]{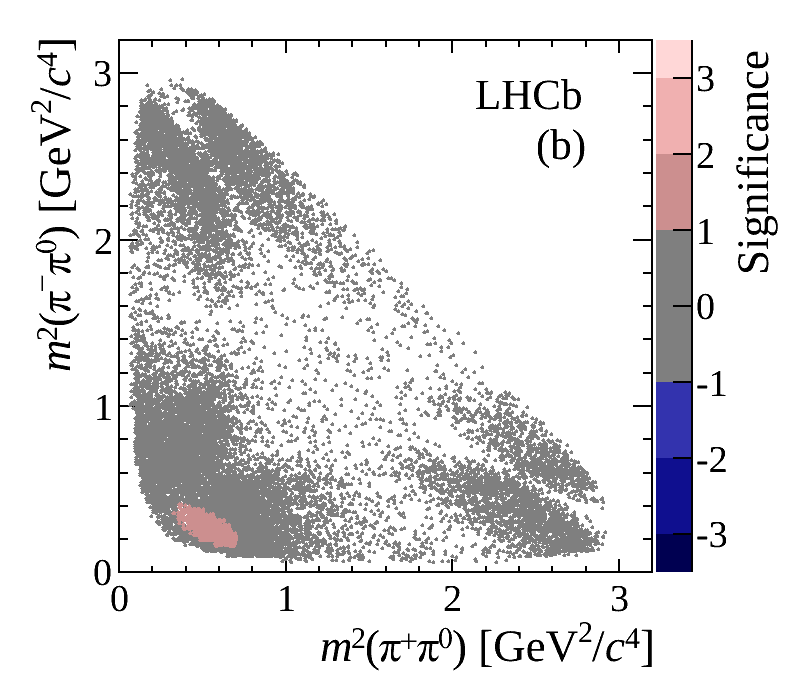}
  \caption{(a) Permutation $T$ value distribution showing the fit function and the measured 
  $T$ value as a red line. (b) Visualisation of local asymmetry significances. 
  The positive (negative) asymmetry significance is set for 
  the \Dz candidates having positive (negative) contribution to the measured $T$ value, respectively (see Sec.\ref{sec:method}).
  \label{fig:results}}
\end{figure}

The application of the energy test to all selected \decay{\Dz}{\pim\pip\piz} candidates using 
a metric parameter of $\sigma=0.3\gevgevcccc$ yields $T=1.84\times10^{-6}$.
The permutation $T$ value distribution is shown in Fig.~\ref{fig:results} (a).
By counting the fraction of permutations with a $T$ value above the nominal $T$ value in the data, 
a \pvalue of $(2.6\pm0.5)\times10^{-2}$ is extracted. 
Alternatively, extrapolation from a fit to the GEV function gives a \pvalue of $(2.1\pm0.3)\times10^{-2}$.
The significance levels of the $T_i$ values are shown in Fig.~\ref{fig:results} (b).
A small phase-space region dominated by the $\rho^+$ resonance contains candidates with 
a local positive asymmetry exceeding $1\sigma$ significance. Varying the metric parameter 
results in the \pvalues listed in Table~\ref{tab:results}; all the \pvalues are at the $10^{-2}$ level. 

\begin{table}[b!]
\caption{Results for various metric parameter values. 
The \pvalues are obtained with the counting method.}
\centering
\begin{tabular}{cc}
\hline \hline 
$\sigma$~[\gevgevcccc] & \pvalue \\ 
\hline  
0.2 & $(4.6 \pm 0.6)\times10^{-2}$ \\
0.3 & $(2.6 \pm 0.5)\times10^{-2}$ \\
0.4 & $(1.7 \pm 0.4)\times10^{-2}$ \\
0.5 & $(2.1 \pm 0.5)\times10^{-2}$ \\
\hline \hline  
\end{tabular}
\label{tab:results}
\end{table}

The data sample has been split according to various criteria to test the stability of the results.
Analyses of sub-samples with opposite magnet polarity, with different trigger configurations, and with fiducial 
selection requirements removing areas of high local asymmetry of the tagging soft pion from the \Dstarp decay 
all show good consistency of the results.

In summary, a search for time-integrated \CP violation in the Cabibbo-suppressed
decay \decay{\Dz}{\pim\pip\piz} is performed using a novel unbinned
model-independent  technique. 
The analysis has the best sensitivity from a single experiment to \CP violation in this decay. 
The data are found to be consistent with the hypothesis of \CP symmetry with a \pvalue of $(2.6\pm0.5)\%$.

\section*{Acknowledgements}

\noindent We express our gratitude to our colleagues in the CERN
accelerator departments for the excellent performance of the LHC. We
thank the technical and administrative staff at the LHCb
institutes. We acknowledge support from CERN and from the national
agencies: CAPES, CNPq, FAPERJ and FINEP (Brazil); NSFC (China);
CNRS/IN2P3 (France); BMBF, DFG, HGF and MPG (Germany); SFI (Ireland); INFN (Italy); 
FOM and NWO (The Netherlands); MNiSW and NCN (Poland); MEN/IFA (Romania); 
MinES and FANO (Russia); MinECo (Spain); SNSF and SER (Switzerland); 
NASU (Ukraine); STFC (United Kingdom); NSF (USA).
The Tier1 computing centres are supported by IN2P3 (France), KIT and BMBF 
(Germany), INFN (Italy), NWO and SURF (The Netherlands), PIC (Spain), GridPP 
(United Kingdom).
This work was supported in part by an allocation of computing time from the Ohio Supercomputer Center.
We are indebted to the communities behind the multiple open 
source software packages on which we depend. We are also thankful for the 
computing resources and the access to software R\&D tools provided by Yandex LLC (Russia).
Individual groups or members have received support from 
EPLANET, Marie Sk\l{}odowska-Curie Actions and ERC (European Union), 
Conseil g\'{e}n\'{e}ral de Haute-Savoie, Labex ENIGMASS and OCEVU, 
R\'{e}gion Auvergne (France), RFBR (Russia), XuntaGal and GENCAT (Spain), Royal Society and Royal
Commission for the Exhibition of 1851 (United Kingdom).

\onecolumn
\addcontentsline{toc}{section}{References}
\setboolean{inbibliography}{true}
\bibliographystyle{LHCb}
\bibliography{main,LHCb-PAPER,LHCb-CONF,LHCb-DP,LHCb-TDR}

\ifx\mcitethebibliography\mciteundefinedmacro
\PackageError{LHCb.bst}{mciteplus.sty has not been loaded}
{This bibstyle requires the use of the mciteplus package.}\fi
\providecommand{\href}[2]{#2}
\begin{mcitethebibliography}{10}
\mciteSetBstSublistMode{n}
\mciteSetBstMaxWidthForm{subitem}{\alph{mcitesubitemcount})}
\mciteSetBstSublistLabelBeginEnd{\mcitemaxwidthsubitemform\space}
{\relax}{\relax}

\bibitem{Grossman:2006jg}
Y.~Grossman, A.~L. Kagan, and Y.~Nir, \ifthenelse{\boolean{articletitles}}{{\it
  {New physics and \CP violation in singly Cabibbo suppressed \PD decays}},
  }{}\href{http://dx.doi.org/10.1103/PhysRevD.75.036008}{Phys.\ Rev.\  {\bf
  D75} (2007) 036008}, \href{http://arxiv.org/abs/hep-ph/0609178}{{\tt
  arXiv:hep-ph/0609178}}\relax
\mciteBstWouldAddEndPuncttrue
\mciteSetBstMidEndSepPunct{\mcitedefaultmidpunct}
{\mcitedefaultendpunct}{\mcitedefaultseppunct}\relax
\EndOfBibitem
\bibitem{LHCb-PAPER-2013-054}
LHCb collaboration, R.~Aaij {\em et~al.},
  \ifthenelse{\boolean{articletitles}}{{\it {Measurements of indirect $CP$
  asymmetries in $D^0\to K^-K^+$ and $D^0\to\pi^-\pi^+$ decays}},
  }{}\href{http://dx.doi.org/10.1103/PhysRevLett.112.041801}{Phys.\ Rev.\
  Lett.\  {\bf 112} (2014) 041801}, \href{http://arxiv.org/abs/1310.7201}{{\tt
  arXiv:1310.7201}}\relax
\mciteBstWouldAddEndPuncttrue
\mciteSetBstMidEndSepPunct{\mcitedefaultmidpunct}
{\mcitedefaultendpunct}{\mcitedefaultseppunct}\relax
\EndOfBibitem
\bibitem{Aubert:2007ii}
BaBar collaboration, B.~Aubert {\em et~al.},
  \ifthenelse{\boolean{articletitles}}{{\it {Measurement of \CP violation
  parameters with a Dalitz plot analysis of
  \decay{B^\pm}{\PD(\pip\pim\piz)\Kpm}}},
  }{}\href{http://dx.doi.org/10.1103/PhysRevLett.99.251801}{Phys.\ Rev.\ Lett.\
   {\bf 99} (2007) 251801}, \href{http://arxiv.org/abs/hep-ex/0703037}{{\tt
  arXiv:hep-ex/0703037}}\relax
\mciteBstWouldAddEndPuncttrue
\mciteSetBstMidEndSepPunct{\mcitedefaultmidpunct}
{\mcitedefaultendpunct}{\mcitedefaultseppunct}\relax
\EndOfBibitem
\bibitem{Grossman:2012eb}
Y.~Grossman, A.~L. Kagan, and J.~Zupan,
  \ifthenelse{\boolean{articletitles}}{{\it {Testing for new physics in singly
  Cabibbo suppressed \PD decays}},
  }{}\href{http://dx.doi.org/10.1103/PhysRevD.85.114036}{Phys.\ Rev.\  {\bf
  D85} (2012) 114036}, \href{http://arxiv.org/abs/1204.3557}{{\tt
  arXiv:1204.3557}}\relax
\mciteBstWouldAddEndPuncttrue
\mciteSetBstMidEndSepPunct{\mcitedefaultmidpunct}
{\mcitedefaultendpunct}{\mcitedefaultseppunct}\relax
\EndOfBibitem
\bibitem{Aubert:2008yd}
BaBar collaboration, B.~Aubert {\em et~al.},
  \ifthenelse{\boolean{articletitles}}{{\it {Search for \CP Violation in
  neutral \PD meson Cabibbo-suppressed three-body decays}},
  }{}\href{http://dx.doi.org/10.1103/PhysRevD.78.051102}{Phys.\ Rev.\  {\bf
  D78} (2008) 051102}, \href{http://arxiv.org/abs/0802.4035}{{\tt
  arXiv:0802.4035}}\relax
\mciteBstWouldAddEndPuncttrue
\mciteSetBstMidEndSepPunct{\mcitedefaultmidpunct}
{\mcitedefaultendpunct}{\mcitedefaultseppunct}\relax
\EndOfBibitem
\bibitem{doi:10.1080/00949650410001661440}
B.~Aslan and G.~Zech, \ifthenelse{\boolean{articletitles}}{{\it New test for
  the multivariate two-sample problem based on the concept of minimum energy},
  }{}\href{http://dx.doi.org/10.1080/00949650410001661440}{J.\ Stat.\ Comput.\
  Simul.\  {\bf 75} (2005) 109}\relax
\mciteBstWouldAddEndPuncttrue
\mciteSetBstMidEndSepPunct{\mcitedefaultmidpunct}
{\mcitedefaultendpunct}{\mcitedefaultseppunct}\relax
\EndOfBibitem
\bibitem{Aslan2005626}
B.~Aslan and G.~Zech, \ifthenelse{\boolean{articletitles}}{{\it Statistical
  energy as a tool for binning-free, multivariate goodness-of-fit tests,
  two-sample comparison and unfolding},
  }{}\href{http://dx.doi.org/10.1016/j.nima.2004.08.071}{Nucl.\ Instrum.\
  Meth.\  {\bf A537} (2005) 626 }\relax
\mciteBstWouldAddEndPuncttrue
\mciteSetBstMidEndSepPunct{\mcitedefaultmidpunct}
{\mcitedefaultendpunct}{\mcitedefaultseppunct}\relax
\EndOfBibitem
\bibitem{Alves:2008zz}
LHCb collaboration, A.~A. Alves~Jr. {\em et~al.},
  \ifthenelse{\boolean{articletitles}}{{\it {The \lhcb detector at the LHC}},
  }{}\href{http://dx.doi.org/10.1088/1748-0221/3/08/S08005}{JINST {\bf 3}
  (2008) S08005}\relax
\mciteBstWouldAddEndPuncttrue
\mciteSetBstMidEndSepPunct{\mcitedefaultmidpunct}
{\mcitedefaultendpunct}{\mcitedefaultseppunct}\relax
\EndOfBibitem
\bibitem{LHCb-DP-2014-002}
R.~Aaij {\em et~al.}, \ifthenelse{\boolean{articletitles}}{{\it {LHCb detector
  performance}}, }{} {LHCb-DP-2014-002}, {in preparation}\relax
\mciteBstWouldAddEndPuncttrue
\mciteSetBstMidEndSepPunct{\mcitedefaultmidpunct}
{\mcitedefaultendpunct}{\mcitedefaultseppunct}\relax
\EndOfBibitem
\bibitem{Breiman}
L.~Breiman, J.~H. Friedman, R.~A. Olshen, and C.~J. Stone, {\em Classification
  and regression trees}, Wadsworth international group, Belmont, California,
  USA, 1984\relax
\mciteBstWouldAddEndPuncttrue
\mciteSetBstMidEndSepPunct{\mcitedefaultmidpunct}
{\mcitedefaultendpunct}{\mcitedefaultseppunct}\relax
\EndOfBibitem
\bibitem{AdaBoost}
R.~E. Schapire and Y.~Freund, \ifthenelse{\boolean{articletitles}}{{\it A
  decision-theoretic generalization of on-line learning and an application to
  boosting}, }{}\href{http://dx.doi.org/10.1006/jcss.1997.1504}{Jour.\ Comp.\
  and Syst.\ Sc.\  {\bf 55} (1997) 119}\relax
\mciteBstWouldAddEndPuncttrue
\mciteSetBstMidEndSepPunct{\mcitedefaultmidpunct}
{\mcitedefaultendpunct}{\mcitedefaultseppunct}\relax
\EndOfBibitem
\bibitem{Pivk:2004ty}
M.~Pivk and F.~R. Le~Diberder, \ifthenelse{\boolean{articletitles}}{{\it
  {sPlot: a statistical tool to unfold data distributions}},
  }{}\href{http://dx.doi.org/10.1016/j.nima.2005.08.106}{Nucl.\ Instrum.\
  Meth.\  {\bf A555} (2005) 356},
  \href{http://arxiv.org/abs/physics/0402083}{{\tt
  arXiv:physics/0402083}}\relax
\mciteBstWouldAddEndPuncttrue
\mciteSetBstMidEndSepPunct{\mcitedefaultmidpunct}
{\mcitedefaultendpunct}{\mcitedefaultseppunct}\relax
\EndOfBibitem
\bibitem{Bediaga:2009tr}
I.~Bediaga {\em et~al.}, \ifthenelse{\boolean{articletitles}}{{\it {On a \CP
  anisotropy measurement in the Dalitz plot}},
  }{}\href{http://dx.doi.org/10.1103/PhysRevD.80.096006}{Phys.\ Rev.\  {\bf
  D80} (2009) 096006}, \href{http://arxiv.org/abs/0905.4233}{{\tt
  arXiv:0905.4233}}\relax
\mciteBstWouldAddEndPuncttrue
\mciteSetBstMidEndSepPunct{\mcitedefaultmidpunct}
{\mcitedefaultendpunct}{\mcitedefaultseppunct}\relax
\EndOfBibitem
\bibitem{Williams:2011cd}
M.~Williams, \ifthenelse{\boolean{articletitles}}{{\it {Observing \CP violation
  in many-body decays}},
  }{}\href{http://dx.doi.org/10.1103/PhysRevD.84.054015}{Phys.\ Rev.\  {\bf
  D84} (2011) 054015}, \href{http://arxiv.org/abs/1105.5338}{{\tt
  arXiv:1105.5338}}\relax
\mciteBstWouldAddEndPuncttrue
\mciteSetBstMidEndSepPunct{\mcitedefaultmidpunct}
{\mcitedefaultendpunct}{\mcitedefaultseppunct}\relax
\EndOfBibitem
\bibitem{Thrust}
{{NVIDIA Corporation,} {\it Thrust quick start guide},
  \href{http://docs.nvidia.com/cuda/pdf/Thrust_Quick_Start_Guide.pdf}{DU-06716-001
  (2014)}}\relax
\mciteBstWouldAddEndPuncttrue
\mciteSetBstMidEndSepPunct{\mcitedefaultmidpunct}
{\mcitedefaultendpunct}{\mcitedefaultseppunct}\relax
\EndOfBibitem
\bibitem{Laura}
T.~Latham {\em et~al.}
\newblock {{\tt Laura++} Dalitz plot fitting package,
  \href{http://laura.hepforge.org/}{\tt http://laura.hepforge.org/}}\relax
\mciteBstWouldAddEndPuncttrue
\mciteSetBstMidEndSepPunct{\mcitedefaultmidpunct}
{\mcitedefaultendpunct}{\mcitedefaultseppunct}\relax
\EndOfBibitem
\bibitem{Sjostrand:2006za}
T.~Sj\"{o}strand, S.~Mrenna, and P.~Skands,
  \ifthenelse{\boolean{articletitles}}{{\it {PYTHIA 6.4 physics and manual}},
  }{}\href{http://dx.doi.org/10.1088/1126-6708/2006/05/026}{JHEP {\bf 05}
  (2006) 026}, \href{http://arxiv.org/abs/hep-ph/0603175}{{\tt
  arXiv:hep-ph/0603175}}\relax
\mciteBstWouldAddEndPuncttrue
\mciteSetBstMidEndSepPunct{\mcitedefaultmidpunct}
{\mcitedefaultendpunct}{\mcitedefaultseppunct}\relax
\EndOfBibitem
\bibitem{Sjostrand:2007gs}
T.~Sj\"{o}strand, S.~Mrenna, and P.~Skands,
  \ifthenelse{\boolean{articletitles}}{{\it {A brief introduction to PYTHIA
  8.1}}, }{}\href{http://dx.doi.org/10.1016/j.cpc.2008.01.036}{Comput.\ Phys.\
  Commun.\  {\bf 178} (2008) 852}, \href{http://arxiv.org/abs/0710.3820}{{\tt
  arXiv:0710.3820}}\relax
\mciteBstWouldAddEndPuncttrue
\mciteSetBstMidEndSepPunct{\mcitedefaultmidpunct}
{\mcitedefaultendpunct}{\mcitedefaultseppunct}\relax
\EndOfBibitem
\bibitem{LHCb-PROC-2010-056}
I.~Belyaev {\em et~al.}, \ifthenelse{\boolean{articletitles}}{{\it {Handling of
  the generation of primary events in \gauss, the \lhcb simulation framework}},
  }{}\href{http://dx.doi.org/10.1109/NSSMIC.2010.5873949}{Nuclear Science
  Symposium Conference Record (NSS/MIC) {\bf IEEE} (2010) 1155}\relax
\mciteBstWouldAddEndPuncttrue
\mciteSetBstMidEndSepPunct{\mcitedefaultmidpunct}
{\mcitedefaultendpunct}{\mcitedefaultseppunct}\relax
\EndOfBibitem
\bibitem{Lange:2001uf}
D.~J. Lange, \ifthenelse{\boolean{articletitles}}{{\it {The EvtGen particle
  decay simulation package}},
  }{}\href{http://dx.doi.org/10.1016/S0168-9002(01)00089-4}{Nucl.\ Instrum.\
  Meth.\  {\bf A462} (2001) 152}\relax
\mciteBstWouldAddEndPuncttrue
\mciteSetBstMidEndSepPunct{\mcitedefaultmidpunct}
{\mcitedefaultendpunct}{\mcitedefaultseppunct}\relax
\EndOfBibitem
\bibitem{Golonka:2005pn}
P.~Golonka and Z.~Was, \ifthenelse{\boolean{articletitles}}{{\it {PHOTOS Monte
  Carlo: a precision tool for QED corrections in $Z$ and $W$ decays}},
  }{}\href{http://dx.doi.org/10.1140/epjc/s2005-02396-4}{Eur.\ Phys.\ J.\  {\bf
  C45} (2006) 97}, \href{http://arxiv.org/abs/hep-ph/0506026}{{\tt
  arXiv:hep-ph/0506026}}\relax
\mciteBstWouldAddEndPuncttrue
\mciteSetBstMidEndSepPunct{\mcitedefaultmidpunct}
{\mcitedefaultendpunct}{\mcitedefaultseppunct}\relax
\EndOfBibitem
\bibitem{Allison:2006ve}
Geant4 collaboration, J.~Allison {\em et~al.},
  \ifthenelse{\boolean{articletitles}}{{\it {Geant4 developments and
  applications}}, }{}\href{http://dx.doi.org/10.1109/TNS.2006.869826}{IEEE
  Trans.\ Nucl.\ Sci.\  {\bf 53} (2006) 270}\relax
\mciteBstWouldAddEndPuncttrue
\mciteSetBstMidEndSepPunct{\mcitedefaultmidpunct}
{\mcitedefaultendpunct}{\mcitedefaultseppunct}\relax
\EndOfBibitem
\bibitem{Agostinelli:2002hh}
Geant4 collaboration, S.~Agostinelli {\em et~al.},
  \ifthenelse{\boolean{articletitles}}{{\it {Geant4: a simulation toolkit}},
  }{}\href{http://dx.doi.org/10.1016/S0168-9002(03)01368-8}{Nucl.\ Instrum.\
  Meth.\  {\bf A506} (2003) 250}\relax
\mciteBstWouldAddEndPuncttrue
\mciteSetBstMidEndSepPunct{\mcitedefaultmidpunct}
{\mcitedefaultendpunct}{\mcitedefaultseppunct}\relax
\EndOfBibitem
\bibitem{LHCb-PROC-2011-006}
M.~Clemencic {\em et~al.}, \ifthenelse{\boolean{articletitles}}{{\it {The \lhcb
  simulation application, \gauss: design, evolution and experience}},
  }{}\href{http://dx.doi.org/10.1088/1742-6596/331/3/032023}{{J.\ Phys.\ Conf.\
  Ser.\ } {\bf 331} (2011) 032023}\relax
\mciteBstWouldAddEndPuncttrue
\mciteSetBstMidEndSepPunct{\mcitedefaultmidpunct}
{\mcitedefaultendpunct}{\mcitedefaultseppunct}\relax
\EndOfBibitem
\end{mcitethebibliography}

\newpage


\newpage
\centerline{\large\bf LHCb collaboration}
\begin{flushleft}
\small
R.~Aaij$^{41}$, 
B.~Adeva$^{37}$, 
M.~Adinolfi$^{46}$, 
A.~Affolder$^{52}$, 
Z.~Ajaltouni$^{5}$, 
S.~Akar$^{6}$, 
J.~Albrecht$^{9}$, 
F.~Alessio$^{38}$, 
M.~Alexander$^{51}$, 
S.~Ali$^{41}$
% ...

{\footnotesize \it
% ...
$ ^{5}$Clermont Universit\'{e}, Universit\'{e} Blaise Pascal, CNRS/IN2P3, LPC, Clermont-Ferrand, France\\
$ ^{6}$CPPM, Aix-Marseille Universit\'{e}, CNRS/IN2P3, Marseille, France\\
% ...
$ ^{9}$Fakult\"{a}t Physik, Technische Universit\"{a}t Dortmund, Dortmund, Germany\\
% ...
$ ^{37}$Universidad de Santiago de Compostela, Santiago de Compostela, Spain\\
$ ^{38}$European Organization for Nuclear Research (CERN), Geneva, Switzerland\\
% ...
$ ^{41}$Nikhef National Institute for Subatomic Physics, Amsterdam, The Netherlands\\
% ...
$ ^{46}$H.H. Wills Physics Laboratory, University of Bristol, Bristol, United Kingdom\\
% ...
$ ^{51}$School of Physics and Astronomy, University of Glasgow, Glasgow, United Kingdom\\
$ ^{52}$Oliver Lodge Laboratory, University of Liverpool, Liverpool, United Kingdom\\
% ...
}
\end{flushleft}

\end{document}